\documentclass[letterpaper]{article} % DO NOT CHANGE THIS
\usepackage{aaai25}  % DO NOT CHANGE THIS
\usepackage{times}  % DO NOT CHANGE THIS
\usepackage{helvet}  % DO NOT CHANGE THIS
\usepackage{courier}  % DO NOT CHANGE THIS
\usepackage[hyphens]{url}  % DO NOT CHANGE THIS
\usepackage{graphicx} % DO NOT CHANGE THIS
\usepackage{natbib}  % DO NOT CHANGE THIS AND DO NOT ADD ANY OPTIONS TO IT
\usepackage{caption} % DO NOT CHANGE THIS AND DO NOT ADD ANY OPTIONS TO IT
\usepackage{algorithm}
\usepackage{algorithmic}
\usepackage{amsmath}
\usepackage{amssymb}
\usepackage{xcolor}
\usepackage{array}

\usepackage{booktabs}
\usepackage{multirow}

\usepackage{newfloat}
\usepackage{listings}
\DeclareCaptionStyle{ruled}{labelfont=normalfont,labelsep=colon,strut=off} % DO NOT CHANGE THIS
\floatstyle{ruled}
\newfloat{listing}{tb}{lst}{}
\floatname{listing}{Listing}
\title{Learning in Multiple Spaces: Few-Shot Network Attack Detection with Metric-Fused Prototypical Networks}
\author {
    Fernando Martinez-Lopez\textsuperscript{\rm 1},
    Lesther Santana\textsuperscript{\rm 1},
    Mohamed Rahouti\textsuperscript{\rm 1}
}
\affiliations {
     \textsuperscript{\rm 1}Computer and Information Science Department, Fordham University, New York, NY 10023, USA\\
      \textbraceleft \textrm{fmartinezlopez, lsantanacarmona, mrahouti}\textbraceright @\textrm{fordham.edu}
}
\usepackage{bibentry}
\begin{document}

\maketitle

\begin{abstract}
Network intrusion detection systems face significant challenges in identifying emerging attack patterns, especially when limited data samples are available. To address this, we propose a novel Multi-Space Prototypical Learning (MSPL) framework tailored for few-shot attack detection. The framework operates across multiple metric spaces—Euclidean, Cosine, Chebyshev, and Wasserstein distances—integrated through a constrained weighting scheme to enhance embedding robustness and improve pattern recognition. By leveraging Polyak-averaged prototype generation, the framework stabilizes the learning process and effectively adapts to rare and zero-day attacks. Additionally, an episodic training paradigm ensures balanced representation across diverse attack classes, enabling robust generalization. Experimental results on benchmark datasets demonstrate that MSPL outperforms traditional approaches in detecting low-profile and novel attack types, establishing it as a robust solution for zero-day attack detection.
\end{abstract}


\section{Introduction} \label{sec:intro}

%\textcolor{red}{Cyber attacks pose a significant and escalating challenge in the realm of cybersecurity, often disrupting organizational operations on a global scale. Conventional detection systems frequently fall short in adapting to the rapidly evolving and diverse nature of these threats, underscoring the critical need for advanced detection frameworks capable of dynamically learning and generalizing across emerging attack patterns, such as zero-day intrusions and rare cyber threats \cite{guo2023review, wei2022abl}.}

% The rapid evolution of cyber threats has become a pressing concern in cybersecurity, mainly as adversaries develop increasingly sophisticated methods to exploit vulnerabilities in critical systems. Traditional network intrusion detection systems (NIDS) have effectively identified known attack patterns. However, these systems struggle to detect emerging, rare, or zero-day attacks, often characterized by limited labeled data and dynamic behaviors \cite{guo2023review, wei2022abl}. This challenge underscores the need for adaptive and data-efficient approaches to enhance threat detection capabilities in real-world scenarios.

The rapid evolution of cyber threats poses a critical challenge in cybersecurity, with adversaries continuously developing sophisticated methods to exploit system vulnerabilities. While traditional network intrusion detection systems (NIDS) excel at identifying known attack patterns, they perform poorly with emerging, rare, or zero-day attacks characterized by limited labeled data and dynamic behaviors \cite{guo2023review, wei2022abl}. This limitation demands adaptive, data-efficient approaches for enhanced threat detection.

% Few-shot learning (FSL) has emerged as a promising solution to address the limitations of traditional methods. By leveraging meta-learning principles, FSL enables models to generalize effectively from minimal labeled data, allowing for the detection of novel attack patterns \cite{wang2020generalizing, duan2021survey}. Early research into FSL applications for cybersecurity demonstrated its potential to classify rare attack types, but issues such as data imbalance, scalability, and generalization to real-world environments persist \cite{yu2020intrusion, yang2022fs}.

Few-shot learning (FSL) addresses these challenges by enabling models to generalize from minimal labeled data through meta-learning principles \cite{wang2020generalizing, duan2021survey}. Early research into FSL applications for cybersecurity demonstrated its potential to classify rare attack types, but issues such as data imbalance, scalability, and generalization to real-world environments persist \cite{yu2020intrusion, yang2022fs}.

Recent advancements have sought to overcome these challenges through innovative frameworks and methodologies. For instance, Ma et al. \cite{ma2023few} introduced a few-shot IoT attack detection framework that integrates adaptive loss weighting to enhance detection performance. Similarly, hybrid and metric-based approaches, such as those proposed by Liang et al., have shown promise in detecting anomalies in complex environments like industrial IoT and cyber-physical systems \cite{liang2021variational, zhou2020siamese}. Despite these advancements, critical gaps still need to be addressed, including the inability to robustly handle diverse attack scenarios and limited scalability for real-world deployment.

% This paper aims to bridge these gaps by introducing a multi-space prototypical learning (MSPL) framework tailored for few-shot attack detection. Building on the success of metric-based learning paradigms, the proposed framework leverages complementary metric spaces, including Euclidean, Cosine, Chebyshev, and Wasserstein distances, to capture diverse topological and distributional properties of attack patterns \cite{tian2021wasserstein, vijayakanthi2021differential}. This integration not only enhances the model's discriminatory power but also improves its robustness in identifying high-profile and low-profile threats.

We present a multi-space prototypical learning (MSPL) framework for few-shot attack detection that leverages complementary metric spaces - Euclidean, Cosine, Chebyshev, and Wasserstein distances - to capture diverse attack pattern properties \cite{tian2021wasserstein, vijayakanthi2021differential}. Building upon Martinez et al.'s work \cite{martinez2024redefining} on dual-space prototypical networks, our framework incorporates Polyak-averaged prototype generation and balanced episodic training to enhance detection stability and representation across attack types.

This paper's key contributions include using Polyak-averaged prototype generation to stabilize the learning process and episodic training paradigms to ensure a balanced representation across diverse attack types. These techniques build upon the foundational work of Martinez et al. \cite{martinez2024redefining}, who demonstrated the efficacy of dual-space prototypical networks for improving detection accuracy in distributed environments. Additionally, the framework is validated on benchmark datasets, showcasing its adaptability and effectiveness in addressing the challenges of data scarcity and emerging cyber threats. The main contributions of this paper are summarized as follows:
\begin{itemize}
    \item By combining multiple metric spaces, the framework enhances the robustness and accuracy of attack detection, addressing the limitations of single-metric models \cite{he2024model}.
    \item The use of Polyak averaging ensures consistent prototype representations, reducing variability during episodic training.
    \item The framework introduces a novel episodic training approach that mitigates the effects of data imbalance, enabling better generalization to unseen attack types \cite{sun2024space}.
    \item Rigorous testing on real-world datasets demonstrates the scalability and effectiveness of the proposed framework in modern cybersecurity applications.
\end{itemize}

% By addressing these challenges, this work not only advances the state of the art in few-shot intrusion detection but also sets the stage for developing more resilient and adaptable cybersecurity systems. The integration of complementary learning paradigms highlights the potential for future research to explore hybrid methodologies that extend the framework's capabilities to multimodal datasets and real-time detection systems \cite{demirpolat2021protedge, ma2023few}.

Tackling these challenges not only progresses the forefront of few-shot intrusion detection but also lays the groundwork for creating more resilient and adaptable cybersecurity systems. The integration of complementary learning paradigms highlights the potential for future research to explore hybrid methodologies that extend the framework's capabilities to multimodal datasets and real-time detection systems \cite{demirpolat2021protedge, ma2023few}.

The rest of this paper is organized as follows. \textbf{Related Work} Section reviews related work on FSL and metric-based approaches for intrusion detection. \textbf{Methodology} Section details the proposed MSPL framework. Next, \textbf{Evaluation} Section presents the results, comparing the framework's performance with state-of-the-art methods. Finally, \textbf{Conclusion} Section concludes with a summary of contributions and potential directions in few-shot attack detection.

\section{Related Work} \label{sec:related}

%This section reviews existing efforts in leveraging FSL and metric-based approaches for intrusion detection, with a particular focus on their applicability to data-scarce environments and complex attack scenarios. 

Given the dynamic and evolving nature of cybersecurity threats, traditional machine learning methods often struggle to generalize across unseen attack types or adapt to new threat vectors. By enabling models to learn from limited labeled data, FSL offers a promising alternative to address these challenges.

\subsection{FSL for Intrusion Detection}

FSL has emerged as a powerful approach for addressing the challenge of learning from limited data, particularly in network intrusion detection \cite{wang2020generalizing, duan2021survey, parnami2022learning}. 
Chowdhury et al. \cite{chowdhury2017few} presented one of the earliest deep learning approaches for FSL in intrusion detection, demonstrating its effectiveness in detecting common attack types despite limited training data. However, their approach faced challenges in scalability and generalization. To enhance the discriminative power of FSL methods, Iliyasu et al. \cite{iliyasu2022few} employed a supervised autoencoder to learn meaningful representations of network traffic. Their approach is particularly effective for low-profile attacks and imbalanced datasets, where traditional methods often fail.

Several other works have applied FSL specifically to network intrusion detection. For instance, \cite{yu2020intrusion} presents an FSL-based intrusion detection framework leveraging prototypical networks, while \cite{yang2022fs} introduces FS-IDS, a framework designed to handle imbalanced training data effectively. In \cite{ma2023few}, a few-shot IoT attack detection method combines self-supervised learning with adaptive loss weighting for enhanced detection performance.

FSL has also been leveraged to address data scarcity in intrusion detection. Aharon et al. \cite{aharon2024few1} introduced a GAN-inspired framework for API attack detection, which generates synthetic samples to enhance training data in environments with limited labeled datasets. This method addresses the challenges of highly dynamic API-based threats and demonstrates the potential of generative approaches in FSL. In a complementary effort, the same authors proposed a classification-by-retrieval framework \cite{aharon2024few2}, which leverages pre-existing embeddings and prototypical representations to detect API anomalies efficiently. This approach provides a practical solution for anomaly detection in real-world systems by emphasizing computational scalability.

Further advancements include methods tailored to specific environments, such as SCADA networks \cite{ouyang2021fs}, industrial IoT \cite{liang2021variational}, and cyber-physical systems \cite{zhou2020siamese}. Other works like \cite{lu2023few} adopt model-agnostic meta-learning to generalize across different intrusion scenarios, and \cite{du2023few} explores class-incremental learning for adapting to evolving attacks. Recently, space-decoupled prototype learning \cite{sun2024space} and graph-based FSL methods \cite{bilot2024few, pan2024few} have been proposed, addressing the limitations of traditional metric-based approaches.

\subsection{Metric-Based Learning in Cybersecurity}

The choice of metric space plays a crucial role in FSL for intrusion detection. Traditional methods rely on Euclidean or cosine metrics for prototype generation \cite{wang2021few}. However, recent works have explored advanced metric spaces to enhance detection performance. For instance, \cite{xu2020method} leverages meta-learning with task-specific adaptations, while \cite{tian2021wasserstein} employs a Wasserstein metric for detecting spoofing attacks in WiFi positioning systems. In \cite{zhou2020siamese}, a Siamese network-based approach integrates multiple metrics for anomaly detection in industrial environments.

Additionally, Miao et al. \cite{miao2023spn} proposed a Siamese prototypical network (SPN) incorporating out-of-distribution detection for traffic classification. This method is particularly adept at identifying anomalous traffic patterns that deviate significantly from known attack profiles, demonstrating the utility of integrating prototypical networks with robust metric-based evaluation mechanisms. Autoencoder-based approaches have also been explored for metric-driven FSL. He et al. \cite{he2021deep} utilized deep autoencoders to capture rich feature representations for malicious traffic detection. Their approach combines feature learning with metric-based classification, achieving high accuracy in few-shot scenarios. Further, Vijayakanthi et al. \cite{vijayakanthi2021differential} introduced a differential metric-based methodology for non-profiled side-channel analysis, primarily in cryptographic contexts. While their work is focused on a different domain, it provides valuable insights into designing adaptive metric-learning systems that could be applied to intrusion detection.

%Lastly, emerging hybrid and ensemble-based techniques are also gaining traction. ProtÉdge \cite{demirpolat2021protedge} combines multiple metrics to improve detection accuracy in software-defined networks, and \cite{rong2021umvd} applies cross-network meta-learning for unseen malware variants. The effectiveness of metric fusion is further demonstrated in \cite{martinez2024redefining}, where a dual-space prototypical network enhances DDoS attack detection by integrating complementary metric spaces.

\subsection{Hybrid and Multi-Space Approaches}

Hybrid and multi-space learning approaches have shown promise in addressing the limitations of single-metric models. For instance, ProtÉdge \cite{demirpolat2021protedge} combines multiple metrics to improve detection accuracy in software-defined networks, and \cite{rong2021umvd} applies cross-network meta-learning for unseen malware variants. The effectiveness of metric fusion is further demonstrated in \cite{martinez2024redefining}, where a dual-space prototypical network enhances DDoS attack detection by integrating complementary metric spaces.

Further studies, such as \cite{iliyasu2022few, kale2023few}, focus on combining supervised and unsupervised learning techniques to improve generalization. Additionally, \cite{he2024model} proposes a model-agnostic framework that integrates generative models with metric-based learning, significantly improving the detection of low-profile attacks. Graph neural networks (GNNs) have also been utilized for few-shot anomaly detection \cite{thein2023few}, demonstrating the potential of graph-based metrics in capturing structural patterns.
 
Qin et al. \cite{qin2020learning} developed a meta-learning-based framework for zero- and few-shot face anti-spoofing. Although their focus is on biometric security, the proposed meta-model offers insights into how meta-learning can generalize across unseen attack scenarios, a principle that could be extended to intrusion detection systems. More efforts include FewM-HGCL \cite{liu2022fewm}, which employs contrastive learning on heterogeneous graphs for malware detection, and \cite{gel2024few}, which utilizes graph contrastive learning for classifying network flow attacks. These approaches underscore the importance of leveraging diverse feature spaces and learning paradigms to tackle the evolving landscape of cyber threats.

%\textcolor{blue}{Further, the integration of generative and metric-based learning frameworks, as seen in recent works, emphasizes the importance of leveraging diverse feature spaces to tackle evolving cybersecurity challenges \cite{aharon2024few1, ma2023few, he2024model}. These efforts highlight the need for adaptive and hybrid approaches to address the complex nature of modern threats.}

\subsection{Uniqueness of this Paper}

This paper introduces the MSPL framework, which uniquely combines four complementary metric spaces—Euclidean, Cosine, Chebyshev, and Wasserstein—through a constrained weighting scheme. Unlike existing single-metric or dual-metric approaches, MSPL leverages the distinct topological properties of multiple distance metrics to enhance detection robustness in data-scarce scenarios. MSPL incorporates two key innovations: Polyak-averaged prototype generation for training stability and a balanced episodic training strategy that ensures equal class representation. These components work together to achieve superior performance in detecting both high-profile and low-profile attacks, particularly excelling at identifying rare and zero-day threats.

\section{Methodology} \label{sec:method}

This section outlines the methodology employed in developing the proposed MSPL framework. It details the design principles, key components, and algorithmic steps that enable the framework to address the challenges of network intrusion detection, such as imbalanced data distributions and the detection of rare attack types.

\subsection{Problem Formulation}
Given a dataset \(\mathcal{D} = \{(\mathbf{x}_i, \mathbf{y}_i)\}_{i=1}^N\) where \(\mathbf{x}_i \in \mathcal{X} \subseteq \mathbb{R}^d\) represents network traffic features and \(\mathbf{y}_i \in \{0,1\}^C\) denotes multi-label intrusion classifications, we propose a multi-space prototypical learning (MSPL) framework. Our objective is to learn a robust embedding function \(f_\theta: \mathcal{X} \rightarrow \mathbb{R}^m\) that maps inputs into a representation space where multiple distance metrics collectively enhance intrusion pattern detection.

\subsection{FSL for Attack Detection}
In the context of network intrusion detection, novel attack patterns often emerge with limited samples. Our framework addresses this FSL scenario through a C-way K-shot formulation, where $C$ represents different attack classes, and $K$ is the limited number of samples per attack type. Formally, for each task:

\[
\mathcal{T} = \{(c, \mathcal{S}_c, \mathcal{Q}_c) | c \in \text{Attack Classes}\},
\]

where \(\mathcal{S}_c\) represents the support set containing $K$ examples of attack class c, and \(\mathcal{Q}_c\) is the query set for evaluation. The stratified sampling ensures the representation of rare attack patterns:

\[
\mathbb{P}(|\mathcal{S}_c| \geq K_{min}) = 1, \forall c \in \text{Attack Classes}
\]

This formulation is particularly crucial for:
\begin{itemize}
\item Zero-day attack detection with limited samples
\item Rapid adaptation to emerging attack patterns
\item Balanced learning across attack types regardless of their frequency
\end{itemize}

\subsection{Multi-Space Prototypical Framework}
Our framework extends traditional prototypical networks through three key innovations:
\begin{enumerate}
    \item Simultaneous operation across complementary metric spaces
    \item Constrained metric space integration
    \item Polyak-averaged prototype generation
\end{enumerate}
%1. Simultaneous operation across complementary metric spaces
%2. Constrained metric space integration
%3. Polyak-averaged prototype generation

The framework operates across four carefully selected metric spaces \(\mathcal{M} = \{E, C, Ch, W\}\), each capturing distinct topological properties:

\subsubsection{Distance metrics:}
For an embedding \(f_\theta(x)\) and prototype \(c_k\), we define:
\begin{enumerate}
    \item Euclidean distance (\(d_E\)): Captures global geometric relationships
\[
d_E(x, c_k) = \|f_\theta(x) - c_k\|_2
\]
    \item  Cosine distance (\(d_C\)): Measures directional similarities
\[
d_C(x, c_k) = 1 - \frac{f_\theta(x) \cdot c_k}{\|f_\theta(x)\| \|c_k\|}
\]
    \item  Chebyshev distance (\(d_{Ch}\)): Identifies maximum deviations
\[
d_{Ch}(x, c_k) = \max_i |f_\theta(x)_i - c_{k_i}|
\]
\item Wasserstein distance (\(d_W\)): Captures distribution-level differences
\[
d_W(x, c_k) = \mathbb{E}_{U \sim \mathcal{U}(0,1)}[|F^{-1}_x(U) - 
    F^{-1}_{c_k}(U)|]
\]
\end{enumerate}

%1. Euclidean distance (\(d_E\)): Captures global geometric relationships
%\[
%d_E(x, c_k) = \|f_\theta(x) - c_k\|_2
%\]

%2. Cosine distance (\(d_C\)): Measures directional similarities
%\[
%d_C(x, c_k) = 1 - \frac{f_\theta(x) \cdot c_k}{\|f_\theta(x)\| \|c_k\|}
%\]

%3. Chebyshev distance (\(d_{Ch}\)): Identifies maximum deviations
%\[
%d_{Ch}(x, c_k) = \max_i |f_\theta(x)_i - c_{k_i}|
%\]

%4. Wasserstein distance (\(d_W\)): Captures distribution-level differences
%\[
%d_W(x, c_k) = \mathbb{E}_{U \sim \mathcal{U}(0,1)}[|F^{-1}_x(U) - 
%    F^{-1}_{c_k}(U)|]
%\]

\subsubsection{Metric space integration:}
Each distance metric undergoes z-score normalization with clipping to ensure comparable scales:
\[
\hat{d}_m(x) = \text{clip}\left(\frac{d_m(x) - \mu_m}{\max(\sigma_m, \epsilon)}, 
    -\gamma, \gamma\right),
\]
where \(\epsilon\) prevents division by zero and \(\gamma\) controls the clipping range.

The integration follows a constrained weighting scheme:
\[
D(x, c_k) = \sum_{m \in \mathcal{M}} w_m \cdot \hat{d}_m(x, c_k)
\]
subject to:
\[
\sum_{m \in \mathcal{M}} w_m = 1, \quad w_m \geq 0 \quad \forall m \in \mathcal{M}.
\]

\subsection{Model Stabilization through Polyak Averaging}

We tested Polyak averaging at the model level to improve stability and convergence in our few-sample setting. Rather than averaging prototypes directly, we maintain an exponential moving average (EMA) of the entire model parameters \(\theta\), which implicitly stabilizes the prototype generation process through a more robust embedding function \(f_\theta\).

For model parameters at training iteration \(t\), the EMA parameters are updated as:
\[
\theta_{\text{EMA}}^{(t)} = \beta\theta_{\text{EMA}}^{(t-1)} + (1-\beta)\theta^{(t)},
\]

where \(\beta\) controls the temporal decay of previous parameter values. This averaging provides three key benefits in our few-shot context: (1) stabilization of the embedding function \(f_\theta\) across episodes, reducing variance in prototype computation, (2) implicit ensembling of models along the optimization trajectory, and (3) smoothing of the optimization landscape to avoid sharp minima that often lead to poor generalization.

The class prototypes \(c_k\) are then computed using the EMA model during inference:
\[
c_k = \frac{1}{|\mathcal{S}_k|} \sum_{(x_i, y_i) \in \mathcal{S}_k} f_{\theta_{\text{EMA}}}(x_i).
\]

\subsection{Episodic Learning Framework}
Our framework implements episodic training through stratified sampling to handle class imbalance in network attacks. The episode construction process begins with initial dataset sampling to ensure class representation:
\[
(\mathcal{X}_{train}, \mathcal{Y}_{train}) = \text{stratified\_sample}(\mathcal{D}, n_{samples}).
\]

Episodes \(\mathcal{E} = \{(\mathcal{S}_i, \mathcal{Q}_i)\}_{i=1}^N\) are then constructed, where each episode contains support and query sets of fixed sizes:
\[
|\mathcal{S}_k| = N_s \text{ and } |\mathcal{Q}_k| = N_q, \forall k \in \{1,\ldots,C\}.
\]

To address the inherent class imbalance in attack detection, we employ an adaptive sampling mechanism. For each class k with insufficient samples, we utilize controlled repetition sampling:
\[
\mathcal{S}_k = \text{sample}(\mathcal{D}_k \times \lceil (N_s + N_q)/|\mathcal{D}_k| \rceil, N_s + N_q).
\]

This sampling strategy maintains episodic learning integrity by ensuring:
\[
\mathcal{S}_i \cap \mathcal{Q}_i = \emptyset, \quad \forall i \in \{1,\ldots,N\}.
\]

The optimization objective employs cross-entropy loss for multi-label classification:

\begin{equation}
\begin{split}
\mathcal{L} = -\sum_{k=1}^C \sum_{x \in \mathcal{Q}} \big(y_k \log(\sigma(-D(x, c_k))) \\ + (1-y_k)\log(1-\sigma(-D(x, c_k)))\big),        
\end{split}
\end{equation}

where \(\sigma\) denotes the sigmoid activation function and \(D(x, c_k)\) represents the combined distance metric across multiple spaces.

\subsection{Training Procedure}
The training process follows Algorithm~\ref{alg:mspl}, which implements our MSPL framework. The algorithm alternates between episode-based training and validation phases, maintaining model stability through gradient clipping and early stopping mechanisms.

For each training iteration, the process:
\begin{enumerate}
    \item Computes embeddings for support and query sets through \(f_\theta\).
    \item Generates class prototypes from support set embeddings.
    \item Calculates and normalizes distances across all metric spaces.
    \item Updates model parameters through gradient descent with clipping.
\end{enumerate}

\begin{algorithm}[tb]
\caption{Multi-space prototypical learning (MSPL).}
\label{alg:mspl}
\textbf{Input}: Dataset \(\mathcal{D}\), episodes \(N_e\), support size \(N_s\), query size \(N_q\), metric weights \(\{w_m\}_{m \in \mathcal{M}}\), Polyak decay \(\beta\)\\
\textbf{Output}: Best model parameters \(\theta^*\)
\begin{algorithmic}[1]
\STATE Initialize model parameters \(\theta\) randomly
\STATE \((\mathcal{X}_{train}, \mathcal{Y}_{train}) \gets\) sample(\(\mathcal{D}\))
\IF{using Polyak averaging}
    \STATE Initialize \(\theta_{\text{EMA}} \gets \theta\)
\ENDIF
\STATE \(\mathcal{E} \gets\) CreateEpisodes(\(\mathcal{X}_{train}, \mathcal{Y}_{train}, N_e, N_s, N_q\))
\FOR{epoch \(= 1\) to \(E\)}
    \FOR{each \((\mathcal{S}, \mathcal{Q}) \in \mathcal{E}\)}
        \STATE \(Z_s \gets f_\theta(\mathcal{S})\)
        \STATE \(Z_q \gets f_\theta(\mathcal{Q})\)
        \STATE \(\{c_k\} \gets\) ComputePrototypes(\(Z_s\))
        \FOR{each metric \(m \in \mathcal{M}\)}
            \STATE \(d_m \gets\) ComputeDistance(\(Z_q, \{c_k\}, m\))
            \STATE \(\hat{d}_m \gets\) NormalizeDistance(\(d_m\))
        \ENDFOR
        \STATE \(D \gets \sum_{m \in \mathcal{M}} w_m \cdot \hat{d}_m\)
        \STATE \(L \gets\) ComputeLoss(\(D, \mathcal{Q}\))
        \STATE Clip gradients and update \(\theta\) using \(\nabla_\theta L\)
        \IF{using Polyak averaging}
            \STATE \(\theta_{\text{EMA}} \gets \beta\theta_{\text{EMA}} + (1-\beta)\theta\)
        \ENDIF
    \ENDFOR
    \STATE Validate model and save if improved
\ENDFOR
\STATE \textbf{return} \(\theta^*\)
\end{algorithmic}
\end{algorithm}
% \vspace{-10pt}

When Polyak averaging is enabled, validation employs the EMA model parameters \(\theta_{\text{EMA}}\), while training proceeds with the primary parameters \(\theta\).

\begin{figure*}[h]
    \centering
    \includegraphics[width=\linewidth]{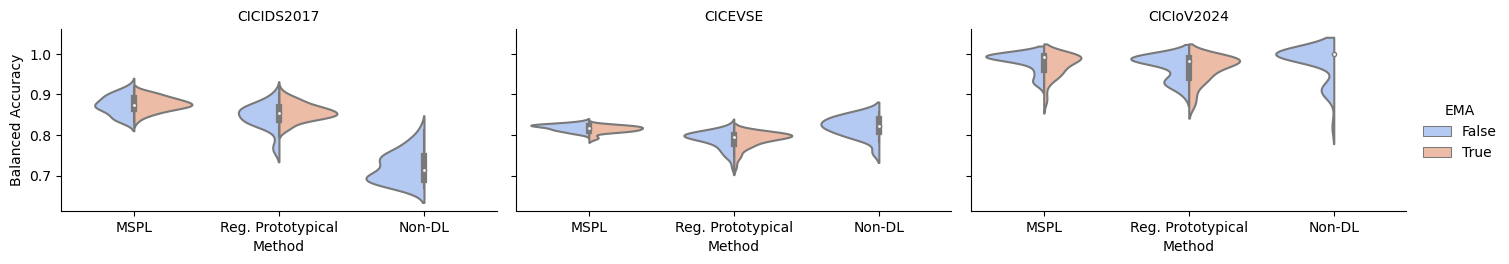}
    \vspace{-15pt}
    \caption{Balanced accuracy comparison for baselines and MSPL approaches across CICEVSE, CICIDS2017, and CICIoV2024 datasets.}
    \label{fig:balanced_accuracy}
\end{figure*}

% \begin{figure}[h]
%     \centering
%     \includegraphics[width=\linewidth]{figures/F1.png}
%     \caption{F1 Score comparison for baseline and multi-space approaches across CICEVSE, CICIDS2017, and CICIoV2024 datasets, with and without Polyak averaging.}
%     \label{fig:F1}
% \end{figure}

\begin{figure*}[h]
    \centering
    \includegraphics[width=\linewidth]{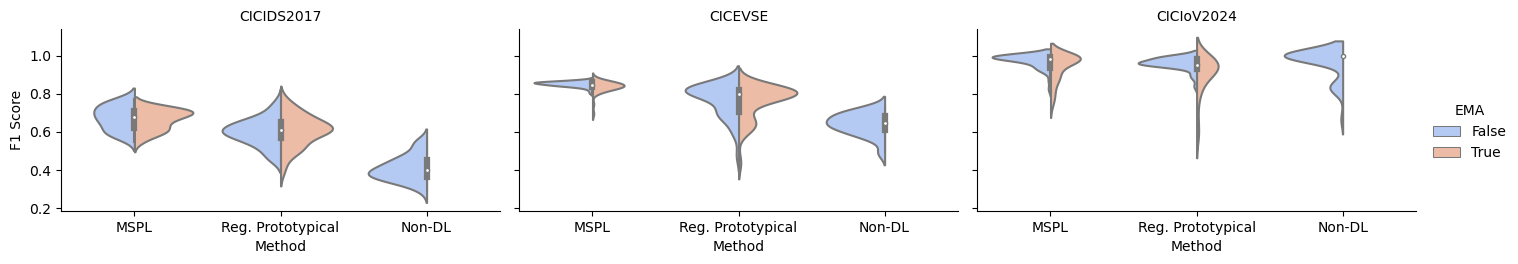}
    \vspace{-15pt}
    \caption{F1-score comparison for baselines and MSPL approaches across CICEVSE, CICIDS2017, and CICIoV2024 datasets}
    \label{fig:F1}
    \vspace{-10pt}
\end{figure*}

% \begin{figure}[h]
%     \centering
%     \includegraphics[width=\linewidth]{figures/AUPRC.png}
%     \caption{AUPRC comparison for baseline and multi-space approaches across CICEVSE, CICIDS2017, and CICIoV2024 datasets, with and without Polyak averaging.}
%     \label{fig:AUPRC}
% \end{figure}

\begin{figure*}[h]
    \centering
    \includegraphics[width=\linewidth]{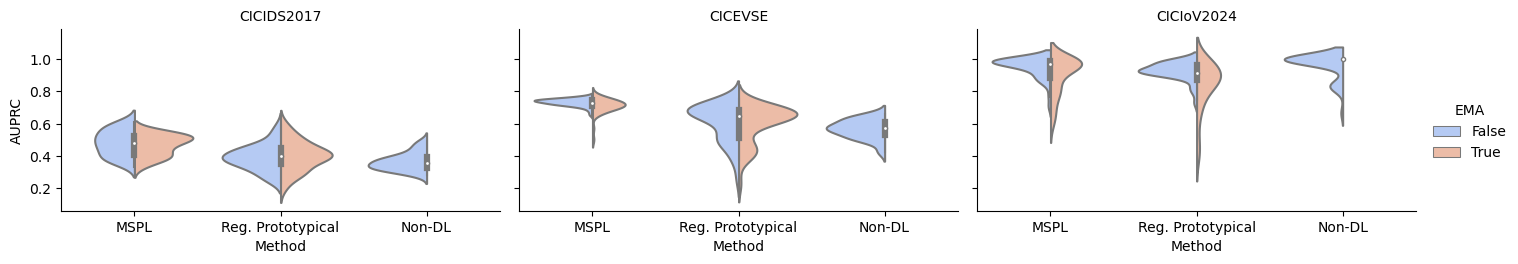}
    \vspace{-15pt}
    \caption{AUPRC comparison for baselines and MSPL experiments across CICEVSE, CICIDS2017, and CICIoV2024 datasets.}
    \label{fig:AUPRC}
\end{figure*}

\section{Evaluation} \label{sec:evaluation}

This section comprehensively evaluates the proposed MSPL framework, demonstrating its efficacy in few-shot network intrusion detection. The experiments employed three benchmark datasets to assess the framework's performance under varied traffic scenarios: CICEVSE2024 \cite{buedi2024enhancing}, CICIDS2017 \cite{sharafaldin2018toward}, and CICIoV2024 \cite{neto2024ciciot}. To rigorously evaluate the FSL capabilities, we restricted the training to only 200 instances. This constrained setting mirrors real-world scenarios where labeled attack data is scarce. The evaluation emphasizes three key metrics: balanced accuracy, validation F1-score, and area under the precision-recall curve (AUPRC), providing insights into the framework's ability to generalize, detect rare attack types, and maintain high classification performance across imbalanced datasets.

\subsection{Performance Analysis}

\subsubsection{CICEVSE2024 dataset:}
On the CICEVSE2024 dataset, the baseline approach achieved a balanced accuracy ranging from 0.7884 to 0.7888 as depicted in Fig. \ref{fig:balanced_accuracy}, with a validation F1-score around 0.76 (Fig. \ref{fig:F1}) and an AUPRC of 0.6081 (Table \ref{combined_table}, Baseline results, and Fig.\ref{fig:AUPRC}). The multi-space approach significantly enhanced these metrics, increasing the balanced accuracy to a range of 0.8143 to 0.8192, raising the validation F1-score to approximately 0.85, and elevating the AUPRC to 0.7333 (Table \ref{combined_table}, Multi-Space results). Integrating tri-metric spaces- Euclidean, Cosine, and Chebyshev- provided complementary perspectives on intrusion patterns, contributing to substantial performance gains.

\begin{table*}[h]
    \centering
    \resizebox{\textwidth}{!}{
    \begin{tabular}{clcccccccccc}
    \toprule
     & Dataset & Polyak & Balanced Accuracy & F1 & AUPRC & \multicolumn{4}{c}{Metric Weights ($w_m$)} \\
    \cmidrule{7-10}
     & & & & & & Euclidean & Chebyshev & Wasserstein & Cosine \\
    \cmidrule{1-10}
    \multirow{3}{*}{\rotatebox{90}{Non-DL}} 
    & CICEVSE & --- & \textbf{0.8212 $\pm$ 0.0079} & 0.6403 $\pm$ 0.0187 & 0.5671 $\pm$ 0.0187 & --- & --- & --- & --- \\
    & CICIDS2017 & --- & 0.7211 $\pm$ 0.0118 & 0.4079 $\pm$ 0.0205 & 0.3647 $\pm$ 0.0170 & --- & --- & --- & --- \\
    & CICIoV2024 & --- & 0.9798 $\pm$ 0.0138 & 0.9623 $\pm$ 0.0257 & \textbf{0.9625 $\pm$ 0.0256} & --- & --- & --- & --- \\
    \midrule
    \multirow{6}{*}{\rotatebox{90}{Baseline}}
    & \multirow{2}{*}{CICEVSE} & $\times$ & 0.7884 $\pm$ 0.0068 & 0.7656 $\pm$ 0.0278 & 0.6081 $\pm$ 0.0377 & 1 & 0 & 0 & 0 \\
    & & $\checkmark$ & 0.7888 $\pm$ 0.0063 & 0.7568 $\pm$ 0.0285 & 0.5957 $\pm$ 0.0374 & 1 & 0 & 0 & 0 \\
    \cmidrule{2-10}
    & \multirow{2}{*}{CICIDS2017} & $\times$ & 0.8460 $\pm$ 0.0101 & 0.5982 $\pm$ 0.0225 & 0.3913 $\pm$ 0.0253 & 1 & 0 & 0 & 0 \\
    & & $\checkmark$ & 0.8572 $\pm$ 0.0066 & 0.6052 $\pm$ 0.0262 & 0.4010 $\pm$ 0.0291 & 1 & 0 & 0 & 0 \\
    \cmidrule{2-10}
    & \multirow{2}{*}{CICIoV2024} & $\times$ & 0.9669 $\pm$ 0.0096 & 0.9591 $\pm$ 0.0106 & 0.9277 $\pm$ 0.0179 & 1 & 0 & 0 & 0 \\
    & & $\checkmark$ & 0.9621 $\pm$ 0.0107 & 0.8999 $\pm$ 0.0338 & 0.8373 $\pm$ 0.0489 & 1 & 0 & 0 & 0 \\
    \midrule
    \multirow{6}{*}{\rotatebox{90}{\textbf{MSPL (Ours)}}} 
    & \multirow{2}{*}{CICEVSE} & $\times$ & 0.8192 $\pm$ 0.0026 & \textbf{0.8508 $\pm$ 0.0046} & \textbf{0.7333 $\pm$ 0.0075} & 1/3 & 1/3 & 0 & 1/3 \\
    & & $\checkmark$ & 0.8143 $\pm$ 0.0029 & 0.8352 $\pm$ 0.0104 & 0.7089 $\pm$ 0.0159 & 1/3 & 1/3 & 0 & 1/3 \\
    \cmidrule{2-10}
    & \multirow{2}{*}{CICIDS2017} 
    & $\times$ & 0.8740 $\pm$ 0.0074 & \textbf{0.6727 $\pm$ 0.0201} & \textbf{0.4798 $\pm$ 0.0254} & 1/3 & 1/3 & 0 & 1/3 \\
    && $\checkmark$ & \textbf{0.8768 $\pm$ 0.0054} & 0.6626 $\pm$ 0.0169 & 0.4658 $\pm$ 0.0208 & 1/2 & 0 & 0 & 1/2 \\
    \cmidrule{2-10}
    & \multirow{2}{*}{CICIoV2024} & $\times$ & \textbf{0.9813 $\pm$ 0.0074} & \textbf{0.9718 $\pm$ 0.0124} & 0.9506 $\pm$ 0.0210 & 1/4 & 1/4 & 1/4 & 1/4 \\
    & & $\checkmark$ & 0.9718 $\pm$ 0.0093 & 0.9404 $\pm$ 0.0228 & 0.8992 $\pm$ 0.0370 & 1/4 & 1/4 & 1/4 & 1/4 \\
    \bottomrule
    \end{tabular}}
    \caption{Performance of MSPL variants under few-shot learning (200 samples) on three network intrusion datasets (40 seeds). Results show mean $\pm$ standard deviation. Non-DL baselines use traditional methods (Logistic Regression, Random Forest, Gradient Boosting). Baseline uses a single-metric (Euclidean) prototypical network. 
    % MSPL explores different metric weights, with $\checkmark$ indicating Polyak averaging.
    Bold values show the best per dataset.}
    \label{combined_table}
    \vspace{-10pt}
\end{table*}

\subsubsection{CICIDS2017 dataset:}
For the CICIDS2017 dataset, the baseline approach exhibited a balanced accuracy between 0.8460 and 0.8572, with modest gains in the validation F1-score and an AUPRC ranging from 0.3913 to 0.4010 (Baseline results in Table \ref{combined_table}, Figures \ref{fig:balanced_accuracy}, \ref{fig:F1}, and \ref{fig:AUPRC}). The multi-space approach improved, achieving balanced accuracy between 0.8740 and 0.8768 and increasing the AUPRC to between 0.4658 and 0.4798 (Table \ref{combined_table}, Multi-Space results). While the validation F1-score exhibited smaller rises than other datasets, the bi-metric space combination of Euclidean and Cosine metrics demonstrated robustness, particularly in handling imbalanced attack scenarios.

\subsubsection{CICIoV2024 dataset:}
The CICIoV2024 dataset displayed high initial performance with the baseline approach, achieving a balanced accuracy between 0.9621 and 0.9669, a validation F1-score ranging from 0.8999 to 0.9591, and an AUPRC between 0.8373 and 0.9277 (Baseline results in Table \ref{combined_table}, Figures \ref{fig:balanced_accuracy}, \ref{fig:F1}, and \ref{fig:AUPRC}). The multi-space approach delivered further enhancements, increasing the balanced accuracy to a range of 0.9718 to 0.9813, advancing the validation F1-score to between 0.9404 and 0.9718, and reaching an AUPRC between 0.8992 and 0.9506 (Table \ref{combined_table}, Multi-Space results). The use of a quad-metric approach- combining Euclidean, Chebyshev, Wasserstein, and Cosine metrics- allowed for balanced
contributions from diverse distance calculations, ensuring comprehensive attack characterization.

\subsubsection{Non-DL Baseline:}
Besides Deep-learning based approaches, we experimented with more traditional machine learning models. On \ref{combined_table} we show the confidence intervals across the benchmark datasets. For CICEVSE the highest performance was achieved using a Random Forest, CICIDS2017 with Gradient Boosting, and for CICIoV2024 using Logistic Regression model. The results show that the traditional approach struggles with more complex datasets like CICIDS2017 and moderately in CICEVSE, where deep learning’s ability to learn hierarchical representations is advantageous. Low F1 and AUPRC values indicate difficulty to capture minority class patterns. However, non-DL performs reasonably well in CICIoV2024, where the structure of the data set might favor simpler algorithms.

\subsection{Polyak Averaging Effectiveness}

The effectiveness of Polyak averaging was evaluated across the datasets, revealing dataset-specific impacts. For the CICEVSE2024 dataset, Polyak averaging introduced minimal variations, with performance metrics remaining stable (Table \ref{combined_table}, Polyak rows). In the case of CICIDS2017, it contributed to marginal stabilization in validation metrics, reflecting modest gains in reliability. The CICIoV2024 dataset exhibited the most significant benefits from Polyak averaging, with noticeable stabilization and smoothing effects (Table \ref{combined_table}, Polyak rows), highlighting its utility in high-performing environments.

\subsection{Metric Space Contribution}

The transition from a single-metric Euclidean baseline to a multi-space approach underlines the importance of metric complementarity. On the CICEVSE2024 dataset, the tri-metric approach enhanced pattern recognition by leveraging distinct geometric and directional properties (Table \ref{combined_table}, Metric Weights column). For CICIDS2017, the bi-metric design effectively reduced overfitting to specific attack types. The quad-metric configuration employed for CICIoV2024 provided the highest adaptability and precision, demonstrating the framework's scalability across diverse scenarios.

\subsection{Key Observations}

%The evaluation revealed several key observations. The MSPL framework demonstrated consistent performance improvements across all datasets, showcasing robust generalization capabilities. Its episodic training and metric fusion approach excelled in limited-sample environments, validating the few-shot learning design. The integration of multiple distance metrics optimized detection accuracy, with each metric contributing unique insights into attack detection. These results emphasize the potential of MSPL as a scalable and effective solution for modern network intrusion detection challenges.

\subsubsection{Generalizability:}
The MSPL framework demonstrated robust generalization capabilities across all three datasets—CICEVSE2024, CICIDS2017, and CICIoV2024. This is evidenced by consistent improvements in balanced accuracy, validation F1-score, and AUPRC compared to the baseline approach. For instance, in the CICEVSE2024 dataset, the balanced accuracy improved from 0.7884–0.7888 (baseline) to 0.8143–0.8192 (multi-space), while the AUPRC increased significantly from 0.6081 to 0.7333. Similarly, for CICIDS2017, the balanced accuracy rose from 0.8460–0.8572 to 0.8740–0.8768, and the AUPRC improved from 0.3913–0.4010 to 0.4658–0.4798. Even on the high-performing CICIoV2024 dataset, where the baseline results were already strong, MSPL achieved further gains in balanced accuracy (0.9621–0.9669 to 0.9718–0.9813) and AUPRC (0.8373–0.9277 to 0.8992–0.9506). These consistent improvements across datasets with different traffic patterns and levels of class imbalance validate the framework’s ability to generalize effectively across diverse scenarios, a critical requirement for intrusion detection systems in real-world settings.

\subsubsection{FSL adaptation:}

The MSPL framework excelled in limited-sample environments, as shown by its strong performance gains, particularly in datasets with lower initial AUPRC values. For example, in the CICIDS2017 dataset, which exhibited lower AUPRC values in the baseline approach (0.3913–0.4010), MSPL demonstrated notable improvements (0.4658–0.4798). The episodic training paradigm, a cornerstone of FSL, likely played a key role in achieving this result by balancing diverse attack representations and ensuring effective adaptation to rare attack types. Moreover, the metric fusion approach, which integrates complementary distance metrics, provided a robust mechanism for learning discriminative embeddings even in the presence of limited data samples. This is particularly evident in the CICEVSE2024 dataset, where the validation F1-score increased from ~0.76 to ~0.85, underscoring the efficacy of MSPL's FSL design in addressing data scarcity and enhancing model adaptability.

\subsubsection{Metric space complementarity:}
The integration of multiple distance metrics was a key factor in optimizing detection accuracy and capturing diverse attack characteristics. Each metric contributed unique insights, as reflected in the framework's performance improvements across datasets. For CICEVSE2024, the tri-metric approach (Euclidean, Cosine, Chebyshev) resulted in performance gains, with the AUPRC increasing from 0.6081 to 0.7333. For CICIDS2017, a bi-metric approach (Euclidean and Cosine) demonstrated effectiveness in mitigating overfitting and improving AUPRC from 0.3913–0.4010 to 0.4658–0.4798. On CICIoV2024, a quad-metric configuration (Euclidean, Chebyshev, Wasserstein, Cosine) provided balanced contributions from all metrics, leading to further improvements in balanced accuracy (0.9718–0.9813) and AUPRC (0.8992–0.9506). These results illustrate the complementarity of the selected metric spaces, as each metric captures different topological or distributional aspects of attack patterns, enabling MSPL to detect both high-profile and low-profile threats effectively.

\section{Conclusion} \label{sec:conclusion}

We proposed a Multi-Space Prototypical Learning (MSPL) framework designed for few-shot attack detection, addressing the challenge of identifying emerging and rare intrusion patterns with limited data samples. By leveraging complementary metric spaces—Euclidean, Cosine, Chebyshev, and Wasserstein distances—our approach captures diverse geometric and statistical properties, enabling comprehensive attack pattern characterization. The integration of Polyak-averaged prototype generation enhances stability and convergence, while the episodic training paradigm ensures balanced class representation and adaptability to diverse attack scenarios.

The framework demonstrated superior performance in detecting zero-day and low-profile attacks, achieving improvements in balanced accuracy and AUPRC compared to traditional methods. Both theoretical and empirical analyses validated the contributions of multi-metric space integration, highlighting its effectiveness and scalability for diverse network intrusion detection tasks. This work provides a foundation for more adaptive and efficient intrusion detection systems, contributing to the ongoing effort to secure modern network environments. Future research will explore the extension of this framework to handle multi-modal datasets and enable real-time detection capabilities.

\bibliography{aaai25}

\end{document}